\documentstyle[12pt]{article}
\makeatletter
\begin{document}
\setlength{\baselineskip}{3ex}
\pagestyle{plain}

\pagenumbering{arabic}
\begin{center}
{\bf \Large Dual Solutions for
Yang-Mills Field Theory in Minkowski Space}
\\
\vspace{1 cm}
{\bf \sf   Liang-Xin Li and Bing-Lin Young}\\
\vspace{1 cm}
Department of Physics and Astronomy, Iowa State
University\\
Ames, Iowa 50011\\
26 March 1994\\
\end{center}
\begin{abstract}
In this paper, We use the spherical symmetrical ansatz to construct a dual
equation
for the non-abelian gauge theory in Minkowski space.The symmetry
in the solution function space is examined.. The analytically
continued instanton satisfies the dual equation but it assumes a
more ansatz. However it is not a solution
of the reduced theory under spherical ansatz  which has the MIT new
solution. This suggests that both
instanton and MIT solutions may be not the lowest energy solution. The
relevant physics is reviewed.
\end{abstract}

\section{Introduction}
It is well known that the classical solutions of non-abelian gauge field
theory have many important physics effect[1],[2]. Especially, the instanton
solution is related  to the tunnelling[3] between different topological
vacuums.
This tunneling can cause explicit violations of conservation laws by the
anomalies[1]. As results of  this, fermion nummber is non-conserved in the
electroweak theory[2], the $\theta$-vacuum arises a strong-CP problem, and the
chirality in QCD is violated[2]. However, all of these effects are understood
by using the instanton as a tunnelling path in Euclidean space[3]. The main
reason for this is that the instanton is only well-behaved and have a well
defined topological identity in Euclidean space.. The analitically
continuation  of
instanton into Minkowski
space  makes the solution very singular. The first semiclassical
calculation[2] shows that the tunnelling probability is expotentially
suppressed and  really small.
However,recently  some group find that when the temperture[5] or energy[6]
becomes very
high,
the field configuration will pass over the barrier instead of tunnelling. This
would make the probability very large. But as you know the instanton is a
solution in the Euclidean space which always corresponds to the tunneling
process. To know what is going on in real physics space time when the energy
becomes bigger than the barrier between the the vacuums, we need to figure out
the actual path in Minkowski space. The classical solution in Minkowski space
will take the role as the instanton in the tunnelling process.

  Since BPST[4] found the instanton solution, there are several new solutions
discovered [7][8][9][10][12]. Recently, MIT group find [11] a new numerical
solution to the equations of motion in Minkowski space by taking the spherical
ansatz and further restrictions. They found that their solution can have
non-integer topological number which leads to the collapse of the periodic
picture of vacuum structure. Moreover, V.V.Khoze[13] has studied the effect of
non-integer topologocal gauge field configurations.

  In this paper, we assume the spherical ansatz[11] [12] and calculate the dual
equations in Minkowski space. We find that there is a local rotation and
translation symmetry
between the parameter function which keep the dual solutions. This symmetry is
shown to the residual gauge transformations which keep the
spherical ansatz. By using the rotational symmetry, one can construct new
solutions. The instanton is analytically continued to Minkowski space. It is
shown that the Minkowski space version of instanton is satisfying the dual
equations. To our surprise is that this solution is not satisfying the MIT
equations which is the equations of motion in the sperical ansatz. However, by
directly pluging the solution into the original field equations, we find it is
a solution of the field equations. In section 2,we introduce the spherical
ansatz and derived the dual equations. The symmetry of the dual solutions is
investigated in section 3. In section 4, we study the solutions of the dual
equations. Conclusions and dicussions are in section 5.

\section{The spherical ansatz and dual equations in the Minkowski space}
In this section, we study the the spherical ansatz[11,12] for $SU(2)$ gauge
group. We follow the notations of MIT's paper. The action for the pure $SU(2)$
Yang -Mills theory is :
\begin{equation}
S=-\frac{1}{2g^2}{\displaystyle\int}d^4xtr(F_{\mu\nu}F^{\mu\nu})
\end{equation}
where $$
F_{\mu\nu}=F_{\mu\nu}^a\frac{\tau^a}{2}=\partial_{\mu}A_{\nu}-\partial_{\nu}A_{\mu}-i[A_{\mu},A_{\nu}]$$
is the field strength and $A_{\mu}=A_{\mu}^a\frac{\tau^a}{2}$. The
space-time metric is given as $\eta_{\mu\nu}=diag(-1,1,1,1)$. $\tau^a$,
$a=1,2,3$ are the Pauli matrices. The spherical
ansatz is given in terms of the four functions $a_0$,$a_1$,$\alpha$,$\beta$
by:
\begin{eqnarray}
A_0(\vec{x},t)&=&\frac{1}{2}a_0(r,t)\vec{\tau}\cdot\hat{x}\nonumber\\
A_i(\vec{x},t)&=&\frac{1}{2}[a_1(r,t)e_i^3+\frac{\alpha(r,t)}{r}e_i^1+\frac{1+
\beta(r,t)}{r}e_i^2]
\end{eqnarray}
where the matrix-valued functions $\{e_i^k\}$ are defined as:
\begin{eqnarray}
e_i^1&=&\tau_i-(\tau\cdot\hat{x})\hat{x_i}\nonumber \\
e_i^2&=&\epsilon_{ijk}\hat{x_j}\tau_k\nonumber\\
e_i^3&=&\tau\cdot\hat{x}\hat{x_i}
\end{eqnarray}
and where $\hat{x}$ is a unit three-vector in the radial direction. Now in
spherical ansatz, the field strength takes the following form:
\begin{eqnarray}
F_{ij}&=&\frac{1}{2}{\displaystyle
\{}\frac{-(\alpha^{\prime}+a_1\beta)}{r}(\tau_i\hat{x}_j-\tau_j\hat{x}_i)+
[\frac{(\beta^{\prime}-a_1\alpha)}{r}+\frac{\alpha^2-2(1+\beta)}{r^2}]
[\nonumber\\
 & &(\hat{x}\times\tau)_j\hat{x}_i-\hat{x}_j(\hat{x}\times\tau)_i]
+\frac{\alpha^2-2(1+\beta)}{r^2}\epsilon_{ijk}\tau_k\nonumber\\
 & &+\frac{(1+\beta)^2}{r^2}\epsilon_{ijl}
\hat{x}_l(\tau\cdot\hat{x}){\displaystyle \}}\\
F_{0i}&=&\frac{1}{2}[\frac{\dot{\alpha}+a_0\beta }{r}e_i^1+\frac{\dot{\beta}-
\alpha
a_0}{r}e_i^2+(\dot{a_1}-{a_0}^{\prime})e_i^3]
\end{eqnarray}
where $\alpha^{\prime}=\frac{\partial\alpha}{\partial r}$, and
$\dot{\alpha}=\frac{\partial\alpha}{\partial t}$,etc.
It is well known that
\begin{equation}
D_{\mu}\tilde{F}^{\mu\nu}=0
\end{equation}
This is the Bianchi identity in gauge theory. So if
\begin{equation}
F^{\mu\nu}=\pm\tilde{F}^{\mu\nu}
\end{equation}
 one can immediately get
\begin{equation}
D_{\mu}F^{\mu\nu}=\pm D_{\mu}\tilde{F}^{\mu\nu}\equiv 0
\end{equation}
which is the equation of motion. So,$F^{\mu\nu}$ is the solution of the field
equation. By using this fact, we require:
\begin{equation}
F^{\mu \nu}=\pm \tilde{F}^{\mu \nu}
\end{equation}
to get the first order differential equations. Now in Minkowski space, the
dual of $F^{\mu\nu}$ is given by:
\begin{equation}
\tilde{F}_{\mu\nu}=\frac{i}{2}{\epsilon_{\mu\nu}}^{\rho\sigma}F_{\rho\sigma}
\end{equation}
(10) becomes $$ F_{\mu\nu}=\pm
\frac{i}{2}{\epsilon_{\mu\nu}}^{\rho\sigma}F_{\rho \sigma}$$
This is equivelent to the following equations
$$ F_{0i}=\pm \frac{i}{2}\epsilon_{ijk}F_{jk}$$
In the parameter of ($\alpha$, $\beta$,$a_0$,$a_1$), the dual equations  is
given as:
\begin{eqnarray}
\dot{\alpha}+a_0\beta&=&\mp i({\beta}^{\prime}-a_1\alpha)\nonumber\\
\dot{\beta}-\alpha a_0&=&\pm i({\alpha}^{\prime}+a_1\beta)\nonumber\\
\dot{a}_1-a^{\prime}_0&=&\pm i\frac{(\alpha^2+\beta^2-1)}{r^2}
\end{eqnarray}
This is from $F_{0i}=\pm \frac{i}{2}\epsilon_{ijk}F_{jk}$. The Euclidean
version of (11) was constructed by E.Witten[12].

\section{symmetries of the dual equations}

To keep the spherical symmmetry, the only residual gauge transformation is as
follows[11]
\begin{equation}
U(\vec{x},t)=exp(i\theta(r,t)\frac{\tau\cdot\hat{x}}{2})
\end{equation}
which is actually a abelian $U(1)$ local transformation. Under this
transformation, the paramters are transformed as
\begin{equation}
A^{\prime}_{\mu}=UA_{\mu}U^{\dagger}+iU\partial_{\mu}U^{\dagger}
\end{equation}
which is
\begin{eqnarray}
a^{\prime}_{\mu}&=&a_{\mu}+\partial_{\mu}\theta(r,t)\nonumber\\
\chi^{\prime}&=&exp(i\theta (r,t))\chi
\end{eqnarray}
where $\chi=\alpha+i\beta$. Under this gauge transformations, the dual
equation(11) is transformed as:
\begin{equation}
F^{\prime}_{\mu\nu}=\pm\tilde{F^{\prime}_{\mu\nu}}
\end{equation}
with $F^{\prime}_{\mu\nu}=UF_{\mu\nu}U^{\dagger}$. So both spherical ansatz
and the dual equations are preserved under(13).

Now suppose we have a group of functions ($\alpha$,$\beta$,$a_0$,$a_1$) which
satisfy equations(11),from(14) or by directly pluging into (11), one can prove
\begin{eqnarray}
\alpha^{\prime}&=&\cos\theta\alpha-\sin\theta\beta\nonumber\\
\beta^{\prime}&=&\sin\theta \alpha+\cos\theta\beta\nonumber\\
a^{\prime}_0&=&a_0+\dot{\theta}(r,t)\nonumber\\
a^{\prime}_1&=&a_1+{\theta}^{\prime}(r,t)
\end{eqnarray}
is also a solution of (11), here $\theta$ is an arbitrary local angle function.
By using this symmetry, one can simplify the dual equation. For instance, we
can work in the $A_0=0$ gauge, and we have a residual rotation with the
$\theta$ function only depending on space coordinates. By using this rotation,
one can further simplify the equation. If we are only interested in the static
solution, we can rotate the $(\alpha,\beta)$ into $(\alpha,0)$ or $(0,\beta)$.
 Furthermore, if ($\alpha$,$\beta$,$a_0$,$a_1$) is the
solution of the dual equations, then ($\alpha$,$\beta$,$-a_0$,$-a_1$) is the,
anti-dual solutions. Through (13), one also can construct new solutions in a
certain gauge.
 Also,we know the confromal
transformation keeps the dual structure[10,11], one can also use those
transformations to make the new solution[11].

\section{The solutions of the spherical dual equations}

Now, by looking at the instanton solution in Euclidean space
\begin{equation}
A_{\mu}=\frac{-ix^2}{x^2+\lambda^2}U\partial_{\mu}U^{\dagger}
\end{equation}
where $x^2=\hat{x}^2+x_0^2$. With the analytical continuation
$x_0\rightarrow it $,$A_0\rightarrow iA_0$. Compared with the ansatz form,
one can get
\begin{eqnarray}
a_0&=&\frac{-2ir}{r^2-t^2+\lambda^2}\nonumber\\
a_1&=&\frac{2it}{r^2-t^2+\lambda^2}\nonumber\\
\alpha&=&\frac{2irt}{r^2-t^2+\lambda^2}\nonumber\\
\beta&=&\frac{r^2+t^2-\lambda^2}{r^2-t^2+\lambda^2}
\end{eqnarray}
By simplily substituting this into the equation(11), one can see that it is the
solution of (11) with the lower sign in the left of (11). We also directly plug
(18) into the field equation to make sure it is a solution of the field
eqaution:
\begin{equation}
D^{\mu}F_{\mu\nu}=\partial^{\mu}F_{\mu\nu}-i[A^{\mu},F_{\mu\nu}]\mbox{$=0$}
\end{equation}

However, to make us suprise, it is not the solution of MIT reduced second
order equation which is supposed to be the field equation in the spherical
ansatz. To look at this, we introduce the MIT methods as follows[11]:

By studying moving spherical shells of energy, MIT group find a new numerical
solution of their reduced field equation. The action in the spgerical ansatz
take the form:
\begin{eqnarray}
S&=&\frac{4\pi}{g^2}{\displaystyle \int}dt{\displaystyle \int}_0^{\infty}dr[-
\frac{1}{4}r^2f_{\mu\nu}f^{\mu\nu}-(D_{\mu}\chi)^{\ast}(D^{\mu}\chi)\nonumber\\
 &  & -\frac{1}{2r^2}(|\chi|^2-1)^2]
\end{eqnarray}
where $f_{\mu\nu}=\partial_{\mu}a_{\nu}-\partial_{\nu}a_{\mu}$,
$\mu$,$\nu$$=(t,r)$,
$\chi=\alpha+i\beta$,$D_{\mu}\chi=(\partial_{\mu}-ia_{\mu})\chi$. The metric
inthe ($1+1$) dimension space is $\bar{\eta}_{\mu\nu}=diag(-1,1)$. From(a),
one can immediately construct the reduced $(1+1)$ dimensional field equations
by $\delta S=0$ which are
\begin{eqnarray}
\partial_{\mu}\frac{\partial\pounds}{\partial(\partial_{\mu}a_{\nu})}&=&\frac{
\partial\pounds}{\partial a_{\nu}}\nonumber\\
\partial_{\mu}\frac{\partial\pounds}{\partial(\partial_{\mu}\chi)}&=&\frac{
\partial\pounds}{\partial\chi}
\end{eqnarray}
where $S={\displaystyle \int}d^4x\pounds$.And the results are the following
second order differential equations[12]:
\begin{eqnarray}
&-\partial^{\mu}(r^2f_{\mu\nu})=i[(D_{\nu}\chi)^{\ast}\chi-{\chi}^{\ast}
D_{\nu}\chi]&\nonumber\\
& [-D^2 + \frac{ 1}{r^2}(|\chi|^2-1) ]\chi=&0
\end{eqnarray}
By solving the (22) numerically, MIT found a new solution which can have
noninteger topological numbers.

 Here, we substitute the solution(18) into (22) and find the solution doesnot
hold the eqautions. This means the equation(22) is not equivalent to the
original field equation. Mathematically, this is understandable: when you take
some certain ansatz of the soultion, it means you put more restictions on the
equation, so the solution for original equations is not necessarily the
solution of the resticted equations. In the Eucledean space, the more
restiction means the higher energy of the solution. If that is the case in the
Minkowski space, the solutions for the equation(22) will have higher energy
than that corresponding solutions of the equations (19). Here, due to the
sigular of the solution (18),we have not prove this yet.

To look at this carefully, we write out (19) explicutly with the
functions($\alpha$,$\beta$, $a_1$, $a_2$). First, $\nu=0$:
\begin{eqnarray}
\frac{(\dot{a}_1-a_o^{\prime})^{\prime}}{2}+\frac{\dot{a}_1-a_0^{\prime}}{r}
+\frac{\beta(\dot{\alpha}+a_0\beta)}{r^2}-\frac{\alpha(\dot{\beta}-a_0\alpha)}{r^2}&=&0
\end{eqnarray}
and $\nu=i$
\begin{eqnarray}
(\alpha^{\prime}+a_1\beta)^{\prime}+a_1(\beta^{\prime}-a_1\alpha)-\frac{\alpha(\alpha^2+\beta^2-1)}{r^2}-\frac{\partial(\dot{\alpha}+a_0\beta)}{\partial
t}-a_0(\dot{\beta}-a_0\alpha)&=0&\nonumber\\
(\beta^{\prime}-a_1\alpha)^{\prime}-\frac{\beta(\beta^2+\alpha^2-1)}{r^2}-\frac{\partial(\dot{\beta}-\alpha
a_0)}{\partial
t}+a_0(\dot{\alpha}+a_0\beta)-a_1(\alpha^{\prime}+a_1\beta)&=0&\nonumber\\
-
-\beta(\alpha^{\prime}+a_1\beta)+\alpha(\beta^{\prime}-a_1\alpha)-\frac{\partial(\dot{a}_1-
a_0^{\prime})}{2\partial t} &=0&
\end{eqnarray}
where we let the respective coefficients of $e_i^j$ equal zero to get the
above equations. Now it is an exercise to check the (11) satisfying the above
equation. Also, we note that when the functions $\alpha$,$\beta$,$a_0$,$a_1$
are real, (23) and the last equation of(24) together can be written as the
first equation of (22),
and the firat+i the second of (24) can be written as the second equation.
However, the solution
(18) is not real.  But the equation is approriate for any case. Nontheless,
the equation (11) means the dual solutions are intrinsically not real but
complex. So the solutions of (11) are not the solutions of (22), however they
are always the solutions of (23) and (24).

 Furthermore, thesolution(18) make the second equation of (11) decoupled. It
is:(18) is satisfying the following equations:
\begin{eqnarray}
\dot{\alpha}+a_0\beta&=&i(\beta^{\prime}-a_1\alpha)\nonumber\\
\dot{\beta}-a_0\alpha&=&0\nonumber\\
\alpha^{\prime}+a_1\beta&=&0\nonumber\\
\dot{a}_1-a^{\prime}_0&=&-i\frac{(\alpha^2+\beta^2-1)}{r^2}
\end{eqnarray}
instead of (11). Obviously, (18) is not the most general solutions of (11).
(18) assumes more restrictions than the spherical ansatz.
\section{Conclusions and Discussions}
By taking the spherical ansatz and assuming the dual relation between fields,
we costruct the dual equations in Minkowski space which claims the analytically
continued instanton as it solutions. The residual symmetry which is a rotation
and translation in
the parameter function space can keep the solutions. The
instanton is a resticted spherical dual solutions which assume more ansatz.
This suggest the instanton may be not the lowest energy solutions. On the
other hand, the solution to equation(11) is not necessarily the solutions of
equation(22) which means the solution of(22) is not the lowest energy solutions
either.
If this is the case, the vacuum structure will have a fruitful picture which
will have a great effect on the anomalies physics.
We will try to find the general solution to equation(11), and investigate the
Minkowski field configurations'effect on the tunnelling process.

\end{document}